\documentclass[%
 11pt,
superscriptaddress,
 amsmath,amssymb,
aps,pre,longbibliography
]{revtex4-1}

\usepackage{setspace}

\usepackage{graphicx}
\usepackage{dcolumn}
\usepackage{bm}
\usepackage{color, soul}
\usepackage{enumitem}
\usepackage{ulem}
\usepackage[utf8]{inputenc}

\usepackage{xcolor}

\usepackage[left]{lineno}

\begin{document}

\title{Leaf-inspired rain-energy harvesting device} 

\author{Jisoo Yuk}
\affiliation{Biological and Environmental Engineering, Cornell University, Ithaca, NY 14853, USA}
\author{Alicia Leem}
\affiliation{Biological and Environmental Engineering, Cornell University, Ithaca, NY 14853, USA}
\author{Kate Thomas}
\affiliation{Mechanical Engineering, Cornell University, Ithaca, NY 14853, USA}
\author{Sunghwan Jung}
\affiliation{Biological and Environmental Engineering, Cornell University, Ithaca, NY 14853, USA}

\date{\today}

\begin{abstract}
We study a rain-powered energy-harvesting device inspired by the natural impact of raindrops on leaves. In nature, a raindrop striking a leaf at high speed causes it to deform and vibrate. Inspired by this, our device uses an elastic beam coupled with a piezoelectric material to convert mechanical vibrations from droplet impacts into electrical energy. We conduct experiments to analyze how beam length, droplet impact location, and residual droplet mass affect energy conversion. Our results, supported by a theoretical model, show strong agreement when the beam length exceeds 5 cm. Beyond this length, the energy conversion becomes independent of further increases, suggesting that 5 cm is optimal for maximizing output. To validate practical applicability, we also test the device under real rain conditions, demonstrating consistent performance. Understanding the interplay between raindrop dynamics and energy conversion can guide the design of efficient, scalable rain-powered energy-harvesting systems for environmental applications.


\end{abstract}


\maketitle 

\section{Introduction} \label{sec:intro} 

A rise in the damaging effects of global warming presents many challenges, including the importance of curtailing the use of fossil fuels to generate energy. Alternatives include renewable energy sources such as solar power \cite{yuksel2008global} and wind \cite{barthelmie2021climate}. However, solar and wind power rely on certain requirements to operate efficiently, and these conditions are not always available. Various regions around the world suffer from a lack of sunshine, but receive constant rainfall. For example, the southeastern United States receives rainfall exceeding 1 mm for over nine months per year, indicating a high potential for rainwater harvesting. The northeastern region is categorized as having medium-high to high potential for rainwater harvesting. In contrast, most of the western United States have the lowest rainwater harvesting potential \cite{loper2019rainwater}. Therefore, discussions are underway on the use of rainfall as a means of obtaining renewable energy under various environmental conditions.

Beyond climatic constraints, solar and wind energy systems also raise ecological concerns. The large land area required for renewable energy power plants can lead to habitat loss, potentially contributing to a decline in biodiversity and disruption of local ecosystems \cite{ashraf2024aligning}. Additional environmental effects include soil erosion, altered hydrology, and localized warming \cite{hernandez2014environmental}. In the case of wind turbines, bird collisions with rotating blades are common and can significantly reduce avian populations \cite{drewitt2008collision}. These disruptions may cascade through food webs, ultimately disturbing ecological balance. As a result, there is a growing need to develop ecologically sensitive designs for renewable energy systems.

Inspired by natural systems, we explore rainfall as a more ecologically harmonious energy source. In nature, many biological bodies are inherently flexible, enabling them to withstand various external forces such as raindrops, snow, wind, and gravity. Along with their flexibility, the unique structural and functional characteristics exhibited by biological organisms serve as important motifs for the development of bio-inspired technologies \cite{schaeffer2025maple, cheng2024weather, park2021fluid}. Among these, we were inspired by the vibrational responses of plant leaves and aimed to identify design principles that enhance energy harvesting performance while remaining compatible with natural environments. For instance, falling raindrops can significantly disturb plant leaves, which undergo deformation under impact at terminal velocities ranging from 2 to 9 m/s \cite{bhosale2020bending}. This deformation includes complex elastic motions, such as bending, twisting, swaying, and localized deflections, facilitated by flexible structures like the petiole \cite{roth2022rain}. 

In addition to deformation, leaves also vibrate due to the momentum transfer from raindrop impacts \cite{lauderbaugh2022biomechanics}. The amount of energy released during these impacts depends on the drop's size, speed, and frequency, which are regulated by the intensity of the rainfall. As rainfall intensity increases, both the number and median size of raindrops increase, up to a saturation point, resulting in greater momentum transfer and impact energy \cite{yakubu2016influence, serio2019raindrop}. Raindrops with diameters ranging from 1 to 5 mm can reach terminal velocities, and the kinetic power they deliver scales nonlinearly with both their diameter and velocity \cite{carollo2018characterizing}. High-resolution lidar and disdrometer studies have further confirmed a strong correlation between terminal velocity and drop diameter, both of which increase with rainfall intensity \cite{aoki2016measurements}. For example, a 4 mm drop may fall at nearly 9 m/s, generating substantial localized forces capable of exciting various vibrational modes in plant tissue. These parameters define the mechanical stress experienced by leaves during natural rainfall, which supports the design of flexible energy harvesters capable of converting dynamic deformations into usable electrical energy.


When rain impacts a beam-like structure, the kinetic energy from the drop is primarily transferred into bending energy. To convert this mechanical energy into electricity, piezoelectric materials are often used. In recent decades, these materials have been widely applied in engineering fields such as sensors, actuators, vibration dampers, motors, and energy harvesters \cite{aydin2023piezoelectric, gripp2018vibration, morita2003miniature, proto2017using, wang2018renewable, mhetre2017human, huang2016design}. Typically configured as thin films with electrodes on both sides, piezoelectric materials generate electric potential in response to mechanical stress. Among many different materials, polyvinylidene fluoride (PVDF) is particularly favored for its flexibility in dynamic environments \cite{lu2020flexible, sukumaran2021recent}. Prior study has modeled leaves as elastic cantilever beams with different surface wettabilities to explore their dynamic response to droplet impacts \cite{gart2015droplet}. Other studies have demonstrated wind-induced energy harvesting using PVDF beams, reporting output powers of 0.855 \textmu W from a PVDF leaf model \cite{wang2018bioinspired} and up to 0.1 mW from a 40 cm² PVDF surface \cite{bhosale2020bending}. While these results show promising potential, the details of mechanical-to-electrical energy conversion and the influence of beam geometry, especially aspect ratio have not been fully characterized.

In this paper, we investigate the energy generation mechanisms of piezoelectric beams of various lengths and their performance under real rain conditions. With a fixed width, a longer beam length correlates to a larger area and more bending, which seems to provide more power intuitively than a shorter beam. However, the interplay between bending dynamics and voltage generation shows saturated energy generation over a certain length of the beam. Here, we demonstrate how beam length affects energy convolution and what relationship there is between the conversion processes of kinetic, bending, and electrical energy.  This interdisciplinary framework bridges biomechanics, environmental physics, and materials science, and provides a foundation for designing bio-inspired rain energy harvesters that are both efficient and ecologically compatible. Through this work, we contribute to the broader effort of developing renewable energy technologies that align with environmental sustainability and the preservation of local ecosystems.

\begin{figure}[t]
\centering
\includegraphics[width=1\textwidth]{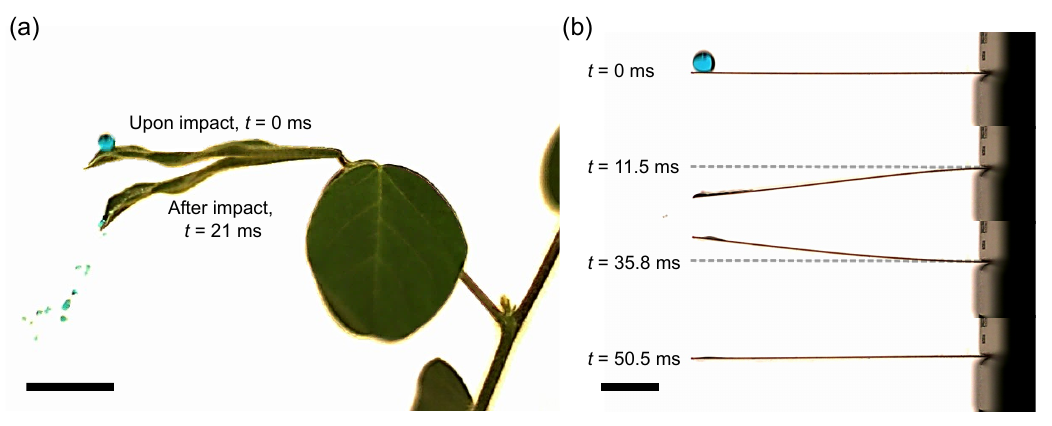}
\caption{ (a) Side view of a water droplet interacting with a soybean leaf. (scale bar = 2 cm) (b) High-speed camera images showing the beam vibration caused by a water droplet impact. (Scale bar = 1 cm)  }
\label{Fig_setup}
\end{figure}

\section{Results}

\subsection*{Beam vibration with a single droplet}

\begin{figure*}[t]
\centering
\includegraphics[width=0.9\textwidth]{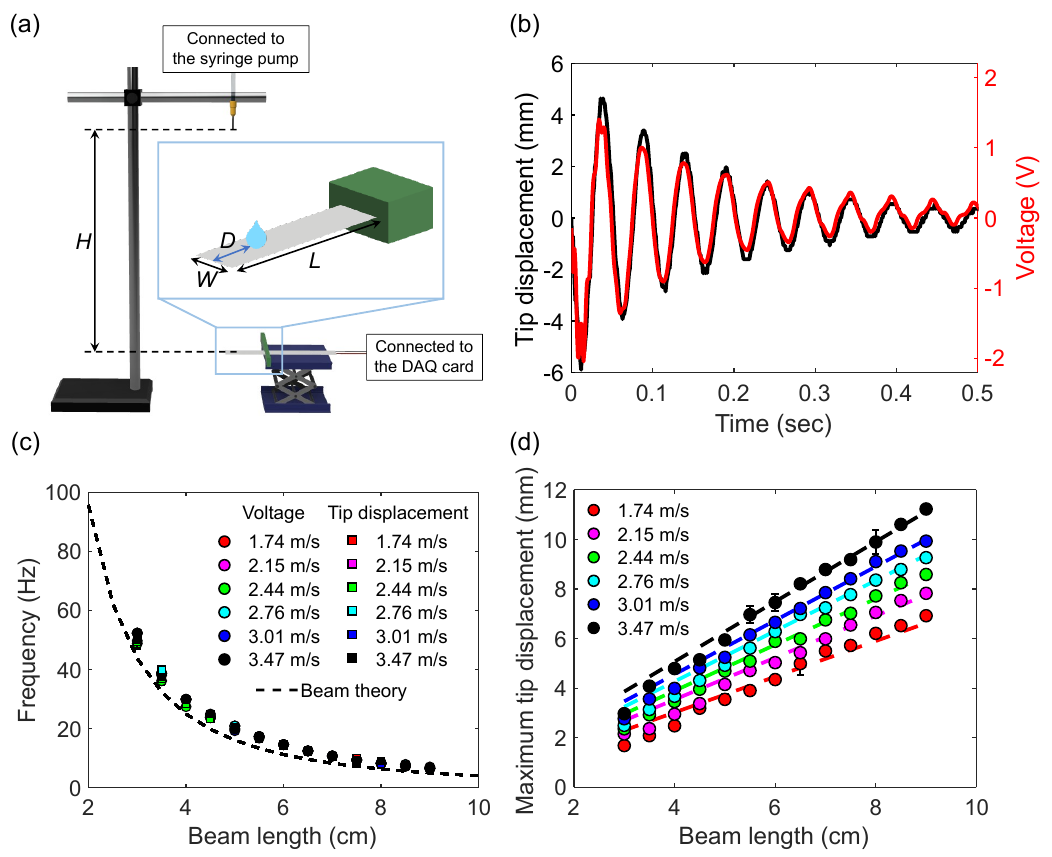}
\caption{(a) Experimental setup where a droplet impacts a piezoelectric beam. The beam has a length  $L$, width $W$, and the droplet impacts the beam at a position, $D$. The vertical distance between the beam and the droplet before impact is denoted as  $H$. (b) Tip displacement and voltage versus time for a 50 mm long beam. (c) Experimental frequency of beam vibration from a single droplet impact as a function of beam length. Circles and squares represent frequencies calculated from voltage and tip displacement, respectively. The black line indicates the theoretical frequency based on beam theory. (d) Maximum tip displacement as a function of beam length. Circles represent experimental data, and the dashed line shows theoretical predictions. Each color in (c) and (d) corresponds to a different droplet impact velocity.}
\label{Fig_char}
\end{figure*}

Figure~\ref{Fig_setup}(a) shows a soybean leaf vibrating in response to a raindrop. The leaf behaves like a beam undergoing vibration. Inspired by this natural phenomenon, Figure~\ref{Fig_setup}(b) presents a 5 cm-long beam that vibrates upon droplet impact at its tip from a height of 60 cm. Figure~\ref{Fig_char}(b) then displays the corresponding tip displacement and voltage response. The vibration at the free end of the beam shows damped harmonic motion. The voltage response closely mirrors the displacement behavior, indicating that beam bending is the primary driver of the piezoelectric effect. In this regard, understanding the mechanics of beam bending is crucial for optimizing the efficiency of electrical energy generation.

The governing equation for the vibration by the droplet impacting on a cantilever beam \cite{erturk2008mechanical, erturk2008distributed, dung2016numerical} can be expressed as
\begin{equation}
\label{goverEq}
EI\frac{\partial^4 y(x,t)}{\partial x^4}
+ c_s I \frac{\partial^5 y(x,t)}{\partial x^4 \partial t}
+ c_a \frac{\partial y(x,t)}{\partial t}
+ m_\mathrm{beam} \frac{\partial^2 y(x,t)}{\partial t^2} = p(t)
\end{equation}
where $E$ is the Young's modulus of the beam, $I$ is the moment of inertia of the cross-section of the beam, $y(x,t)$ is the tip displacement of the cantilever beam, $c_s$ is the coefficient of strain rate damping,  $c_a$ is the viscous air damping coefficient, $m_{beam}$ is the mass per unit length of the cantilever beam, and $p(t)$ is the external excitation. In our experiment, $p(t)$ represents a drop impact described by the Dirac delta function that can be written as
\begin{equation}
\label{dirac}
p(t) = J \delta(x - x_p) \delta(t - \tau)
\end{equation}
where $J$ is the impulse, $x_p$ is the impact location, and $\tau$ is the moment of impact. 


The tip displacement, $y(x,t)$ can be separated into eigenfunctions of space and time, and is expressed as
\begin{equation}
\label{defl}
y(x,t) = \sum_{r=1}^{\infty} \phi_r(x) \eta_r(t)
\end{equation}
where $\phi_r(x)$ is the mass-normalized eigenfunction in space and $\eta_r(t)$ is the modal response in time, and $r$ is the mode number. By separating the eigenfunctions of space and time, we can obtain individual solutions for $\phi_r(x)$ and $\eta_r(t)$. By combining them, we then obtained the theoretical solution for the beam’s displacement. Further details are provided in the Supplementary Information Section A.

When the drop impacts the tip of the beam, the first mode contributes the most significantly. If we simplify the equation by focusing only on the first mode and normalize the mode shape such that $\phi_1(L)=1$, the displacement $y(x,t)$ can be expressed as 
\begin{equation}
\label{defl_sol}
    y(x,t)= \phi_1(x)\eta_1(t) \approx -\Delta_\mathrm{M}e^{-\omega_1\zeta_1 t}\cos(t\omega_1)
\end{equation}
where $\Delta_M$ is the maximum displacement, and $\omega_1$ is the natural frequency of the 1st mode vibration expressed as \cite{meirovitch2010fundamentals, gart2015droplet}
\begin{equation}
\label{omega_d}
\omega_1=\beta_1^2\sqrt{\frac{EI}{m_\mathrm{beam}L+M_\mathrm{drop}}}\frac{1}{L^{3/2}}
\end{equation}
with $\beta=1.875$. Since the beam initially bends downward, a negative sign is applied in Equation~\eqref{defl_sol}.

\subsection*{Frequency of the beam}

Figure~\ref{Fig_char}(c) shows the experimental and theoretical frequencies of beams ranging from 3 to 9 cm at six different velocities. To calculate the experimental frequencies of the tip displacement and voltage signals, each curve was fitted to a damped harmonic oscillation using the least-squares method. The frequencies calculated from both the tip displacement and voltage signals were identical for each beam length, and the frequency differences due to the impact velocity of the droplet could be negligible. The frequency decreases as the length of the beam increases.

The theoretical frequency value is calculated using Equation~\eqref{omega_d} for the first mode. In this equation, if the $M_\mathrm{drop}/m_\mathrm{beam}$ is defined as $\chi$ in the unit of length, and the non-dimensional length ratio, $\chi/L$, is much smaller than 1 (experimentally measured between 0.045 and 0.14), the frequency is proportional to $L^{-2}$ like Equation S5. 

Compared to the theoretical predictions, both the tip displacement frequency and the voltage frequency shows similar trends. This indicates that the experimental frequencies, like the theoretical ones, depend on the beam length.

However, the experimental frequencies were slightly higher than the theoretical values. Such underestimations in theoretical predictions of droplet-induced beam vibrations have also been reported in previous studies \cite{fowler2024resonance}. This may be due to additional nonlinear effects arising from the beam’s flexibility.

\subsection*{Maximum tip displacement by droplet impact}

Figure~\ref{Fig_char}(d) shows the maximum tip displacement of beams ranging from 3 to 9 cm in length at six different impact velocities. The experimental values were determined from the peak amplitude of the damped harmonic oscillation. Overall, the results indicate that the tip displacement increases with both beam length and droplet velocity.

Theoretically, by combining Equations (3, 4, S1, S7, S8) for the first vibration mode, we find that when $\chi/L \ll 1$, the maximum tip displacement $\Delta_\mathrm{M}$ scales linearly with beam length ($\Delta_\mathrm{M} \propto L$). The displacement is also influenced by the impact velocity of the droplet, which contributes to the impulse. Since the impulse occurs over a short time scale, it can be approximated as $J \approx M_\mathrm{drop}v/\gamma$, where $v$ is the impact velocity and $\gamma$ is a time-scale factor related to droplet spreading (see Supplementary Information Section B). As a result, it leads to a linear relationship between displacement and velocity, as confirmed experimentally (Figure S1).

The dashed lines in Figure~\ref{Fig_char}(d) represent the theoretical predictions, which align well with the experimental data for beams longer than 5 cm. However, due to the small $\chi/L$ assumption, the theory tends to overestimate displacements for shorter beams.



\subsection*{Damping ratio}
Figure S2 presents the damping ratio of the voltage and tip displacement for beams ranging from 3 to 9 cm in length. The experimental damping ratio was determined by comparing the first and fifth peaks of the voltage and tip displacement signals to calculate the logarithmic decrement and, subsequently, the damping ratio. The damping ratios obtained from both tip displacement and voltage measurements ranged from 0.048 to 0.05 for beam lengths between 4 and 9 cm, with the exception of the 3 cm beam. This indicates that for beam lengths between 4 and 9 cm, the damping ratio exhibits no significant length dependence.

\subsection*{Energy generation}

\begin{figure*}[t]
\centering
\includegraphics[width=0.9\textwidth]{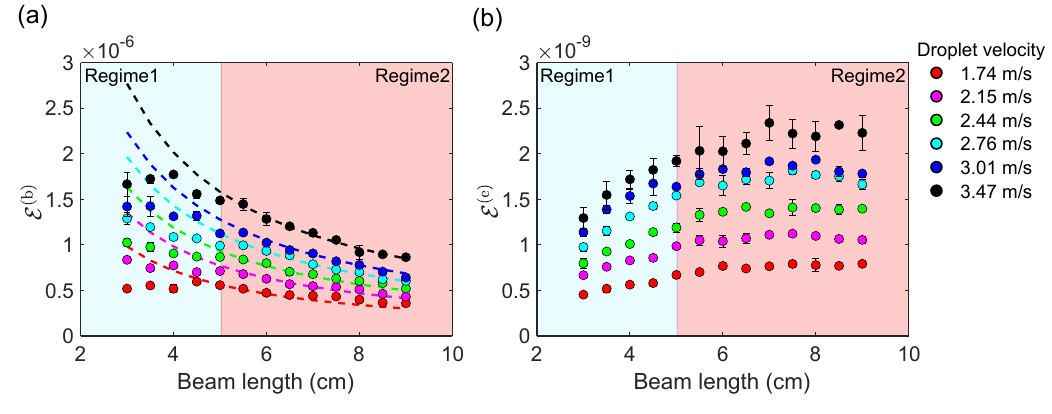}
\caption{(a) Time-averaged bending energy and (b) time-averaged electrical energy as a function of beam length. Circles represent experimental data, and dashed lines indicate linear fits for each velocity condition. Each color in the figure corresponds to a different droplet impact velocity.}
\label{Fig_2}
\end{figure*}

In raindrop energy harvesting, the kinetic energy of the droplet is first converted into the bending energy of the beam, which is then transformed into electrical energy. We describe the different forms of energy generated during raindrop energy harvesting and investigate how these energies are related to the beam length.

First, the kinetic energy ($\mathcal{E}^\mathrm{(k)}$) of the droplet is given by 
\begin{equation}
\label{BE_ori}
\mathcal{E}^\mathrm{(k)} = \frac{1}{2} M_\mathrm{drop} v^2
\end{equation}
which is independent of beam length.

Second, the bending energy ($\mathbb{E}^\mathrm{(b)}$) is given by $\mathbb{E}^\mathrm{(b)} = \frac{1}{2} P y(t)$, where $P = \frac{3EIy(t)}{L^3}$. The time-averaged bending energy ($\mathcal{E}^\mathrm{(b)}$) of the piezoelectric beam is 

\begin{equation}
\label{BE_avg}
    \mathcal{E}^\mathrm{(b)} = \frac{1}{t_f-t_0} \int_{t_0}^{t_f} \frac{3EI}{2} \frac{y(t)^2}{L^3}dt
\end{equation}
where $t_0$ is the initial time just before the vibration begins, and $t_f$ is the time after ten vibrations. Solving this equation \eqref{BE_ori} with equation \eqref{defl_sol}, we obtain

\begin{equation}
\label{BE}
    \mathcal{E}^\mathrm{(b)} = \frac{3{\Delta_\mathrm{M}}^2EI(3\zeta_1^2+1)}{8t_f\omega_1\zeta_1 (\zeta_1^2+1)L^3}
\end{equation}


In equation \eqref{BE}, $\zeta_1$, $EI$, and $t_f\omega_1$ are independent of length, while the length-dependent terms are ${\Delta_\mathrm{M}}^2$ on the numerator and $L^3$ on the denominator. Previously, we established that when $\chi/L \ll 1$, $\Delta_\mathrm{M}$ is linearly proportional to $v$ and $L$. Consequently, the product of ${\Delta_\mathrm{M}}^2$ and $L^{-3}$ results in $\mathcal{E}^\mathrm{(b)} \propto v^2L^{-1}$.

Figure~\ref{Fig_2}(a) shows the time-averaged bending energy and fitted lines for six different impact velocities to check whether the experimental data reflected what we theoretically expected. For beams longer than 5 cm, inverse fitting at each velocity (1.74–3.47 m/s) yields $R$-squared values of 0.52, 0.88, 0.90, 0.89, 0.89, and 0.97, respectively, confirming that our theoretical prediction is well-supported for longer beams. However, for shorter beams under 5 cm ($\chi/L > 0.081$), data points deviate from the expected trend. This suggests that the trend does not hold well for shorter beams, likely because the criterion assumes $\chi/L \ll 1$. To clarify the observed tendencies, we categorized the beams into two regimes: Regime 1 (short beams, $L<$5 cm) and Regime 2 (long beams, $L\geq$5 cm).

Lastly, the electrical energy ($\mathbb{E}^\mathrm{(e)}$) is given by $\mathbb{E}^\mathrm{(e)} = \frac{1}{2} C V^2$. The time-averaged electrical energy ($\mathcal{E}^\mathrm{(e)}$) of the piezoelectric beam is

\begin{equation}
\label{EE}
    \mathcal{E}^\mathrm{(e)} = \frac{1}{t_f-t_0} \int_{t_0}^{t_f} \frac{CV^2}{2}dt
\end{equation}
where $C =11$ nF is the capacitance of the piezoelectric beam that we used in this study, and $V$ is the voltage. Figure~\ref{Fig_2}(b) presents the time-averaged electrical energy as a function of beam length for six different velocities. The results indicate that beam length influences electrical energy generation. In Regime 1, time-averaged electrical energy increases with beam length. However, in Regime 2, the trend plateaus, showing no further increase in electrical energy. In addition, higher impact velocities result in greater electrical energy generation.

\subsection*{Energy conversion process}
We were able to establish the relationships using the respective energy equations, which are discussed in more detail in Supplementary Information Section C. By integrating the two relationship equations, one from kinetic energy to bending energy ($\mathcal{E}^\mathrm{(b)}\propto WL^{-1}\mathcal{E}^\mathrm{(k)}$) and the other from bending energy to electrical energy ($\mathcal{E}^\mathrm{(e)}\propto WLW^{-2}\mathcal{E}^\mathrm{(b)}$), we derive the final relationship equation from kinetic energy to electrical energy: 
\begin{equation}
\mathcal{E}^\mathrm{(e)}\propto \mathcal{E}^\mathrm{(k)}
\end{equation}

\begin{figure*}[t]
\centering
\includegraphics[width=1\textwidth]{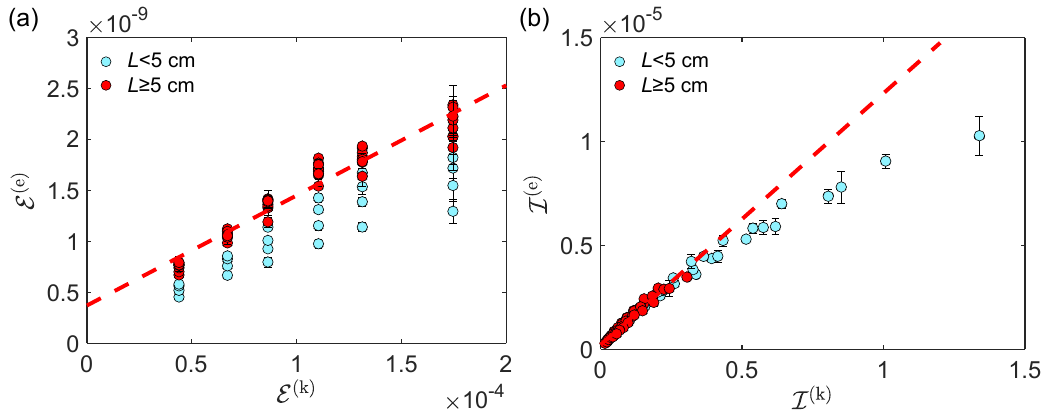}
\caption{(a) Kinetic energy versus electrical energy, and (b) kinetic energy intensity versus electrical energy intensity. Sky-blue symbols represent data from beams shorter than 5 cm, while red symbols correspond to beams that are 5 cm or longer. The red dashed line indicates a fitted curve based on the red data points, with $R$-squared values of 0.94 for (a) and 0.97 for (b), respectively.}
\label{Fig_3}
\end{figure*}

In this process, all terms related to length were eliminated, effectively eliminating the dependency on length in the energy conversion from kinetic energy to electrical energy. Since all of these assumptions rely on $\chi/L \ll 1$, the relationship holds only for sufficiently long beams, specifically those with a length of 5 cm or more. According to this equation, when $L \geq 5$ cm, the generation of $\mathcal{E}^\mathrm{(e)}$ at the same $\mathcal{E}^\mathrm{(k)}$ becomes saturated, indicating that $L = 5$ cm is the optimal beam length for maximizing electrical energy generation. Figure~\ref{Fig_3}(a) shows that beams longer than 5 cm tend to collapse onto a single curve, which supports the validity of the theoretical prediction.


To further analyze the energy conversion, we normalize the energy by defining energy intensity ($\mathcal{I}$), dividing the energy by area and time to account for spatial and temporal variations. The reason for introducing this density term is that, in the context of rainfall, an extended version of a single droplet impact, energy is typically characterized per unit area and time (unit=$\mathrm{J/(m^2hr)}$).

\begin{equation}
\label{normBE}
    \mathcal{I}^\mathrm{(i)} = \frac{\mathcal{E}^\mathrm{(i)}}{\mathrm{Area}\times \mathrm{Time}}
\end{equation}

Figures S3(b) and S4(c) show the relationships of $\mathcal{I}^\mathrm{(b)}$ to $\mathcal{I}^\mathrm{(k)}$ and $\mathcal{I}^\mathrm{(e)}$ to $\mathcal{I}^\mathrm{(b)}$, respectively. In both plots, the data collapse when $L \geq 5$ cm. Similarly, in Figure~\ref{Fig_3}(b), the final relationship, $\mathcal{I}^\mathrm{(e)}$ to $\mathcal{I}^\mathrm{(k)}$, collapses into a single linear trend with an $R$-squared value of 0.97 when $L \geq 5$ cm. This confirms that electrical energy and kinetic energy exhibit no length dependency in Regime 2.

\subsection*{Effect of the droplet's impact location}

\begin{figure*}[t]
\centering
\includegraphics[width=0.9\textwidth]{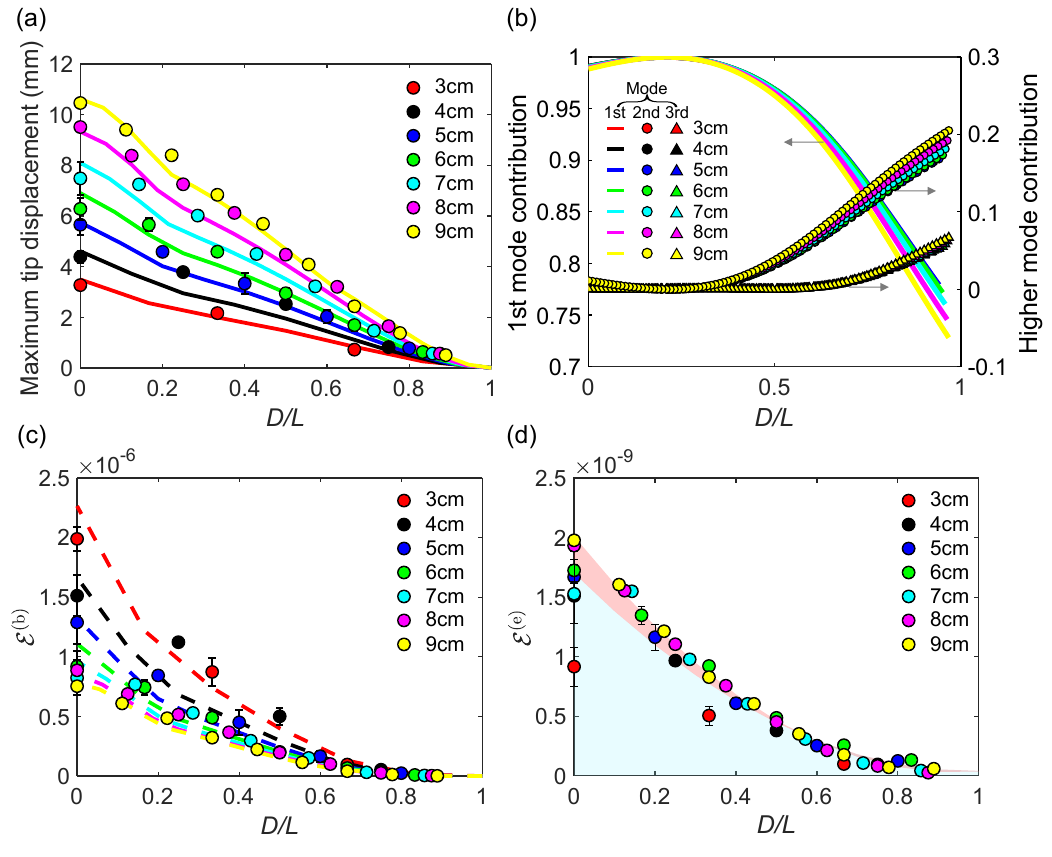}
\caption{Experimental test data showing the effect of droplet deposition location. (a) Maximum tip displacement depending on the drop impact location. The drop position, $D$, is normalized by the beam length, $L$, expressed as $D/L$. The circular markers represent experimental data, and the solid lines of the same color indicate theoretical results. Both experimental and theoretical trends show that the maximum tip displacement decreases as $D/L$ approaches 1. (b) Contribution of the first,  second, and third vibration modes depending on the drop location. The solid lines represent the contribution of the first mode, while the circular and triangular markers show the contribution of the second and third mode, respectively. Up to $D/L$ = 0.25, the first mode dominates with over 98\% contribution, but beyond $D/L$ = 0.25, the second mode gradually becomes more significant. Also, beyond $D/L = 0.6$, a minor contribution from the third mode is observed. (c) Time-averaged bending and (d) time-averaged electrical energy as a function of the droplet deposition location ratio $D/L$. Each color in (c) and (d) represents the beam length. Dashed line in (c) shows the theoretical bending energy.}
\label{Fig_4}
\end{figure*}

After examining the beam’s response when a single droplet landed on its tip, we expanded our investigation to analyze how the beam reacts when a droplet impacts various locations along its length and how this influences energy conversion.

First, we analyzed how the frequency, and $\Delta_\mathrm{M}$ vary with the droplet’s landing position. When the droplet landed closer to the clamped fixed point, the beam tended to exhibit higher mode vibrations, resulting in a faster frequency (Figure S5). Specifically, the frequency increased by about 10\%, though this effect diminished for beams longer than 5 cm.

Next, we investigated the tip displacement affected by the droplet’s landing location. When a droplet strikes a position other than the tip, it excites higher modes of vibration, which can influence the overall vibration pattern. To account for these higher mode effects, we solved equation~\eqref{defl} by incorporating the first, second, and third modes, which are the most influential in this context. To accurately reflect the droplet’s landing position in our mathematical model, we adjusted the external force term, $p(t)$, to include the exact impact location with $x_p=L-D$, where $D$ represents the distance of the droplet from the tip. We also considered the effect of the droplet’s mass on the beam's displacement. When the droplet impacts near the tip, a portion of the droplet mass is typically shed immediately after the collision, resulting in minimal contribution to the beam's deflection from the droplet’s weight. In contrast, when the droplet lands closer to the base, more of its mass tends to remain attached to the beam, leading to greater tip displacement observed in experiments.

Figure S6 shows the tip displacement and voltage response of 5 cm beam, when the droplet impacted $D=0$ mm and $D=40$ mm. When the droplet impacts at $D=40$ mm, the tip displacement and overall response are significantly reduced compared to the case of $D=0$ mm. The location of droplet impact is a key factor in determining the amplitude of the beam's vibration. Figure~\ref{Fig_4}(a) shows how $\Delta_\mathrm{M}$ change as the droplet landing position varies with 3 to 9 cm beam. To represent the droplet position relative to the beam length, the horizontal axis is expressed as $D/L$. Both the experimental and theoretical values indicate that the highest $\Delta_\mathrm{M}$ occurs when the droplet impacts at the tip, gradually decreasing as the impact point moves closer to the clamped end. This is because the closer the impact is to the clamped end, the more the impact energy is distributed across multiple vibrational modes, and the shorter the effective moment arm becomes, limiting the overall deflection. Generally, the theoretical values align well with the experimental data. 

In the theoretical calculations, we were able to evaluate the contributions of the first, second, and third modes to the actual $\Delta_\mathrm{M}$. The results showed that the influence of the second and third mode becomes significant beyond $D/L=0.25$ and $D/L=0.6$, respectively (Figure~\ref{Fig_4}(b)).

Using the theoretically obtained $\Delta_\mathrm{M}$ values, we were able to calculate the bending energy by applying Equation\eqref{BE}. As shown by the dashed lines in Figure~\ref{Fig_4}(c), the theoretical trend closely follows the experimental data. Experimentally, we also determined the corresponding electrical energy. When normalized by $D/L$, the electrical energy data for beams longer than 5 cm collapses within the red shaded region (Figure~\ref{Fig_4}(d)). This indicates that, regardless of the drop location, electrical energy generation tends to saturate beyond a beam length of 5 cm when subjected to the same kinetic energy.

\subsection*{Effect of multiple impacting droplets}
We also investigated the scenario in which multiple droplets continuously impacted the same location. The experiment was conducted simply by dropping droplets sequentially at the same position, and the changes in bending and electrical energy were measured. Both experimental and theoretical analyses confirmed that residual mass plays a role in reducing the vibration frequency and $\Delta_\mathrm{M}$ values, which ultimately contributes to the reduction of bending energy and electrical energy generation (see Figure S7). Detailed equations and illustrations can be found in Figure S7, S8 and Supplementary Information Section D.

\subsection*{Field tests under natural rainfall}

\begin{figure*}[htbp!]
\centering
\includegraphics[width=1\textwidth]{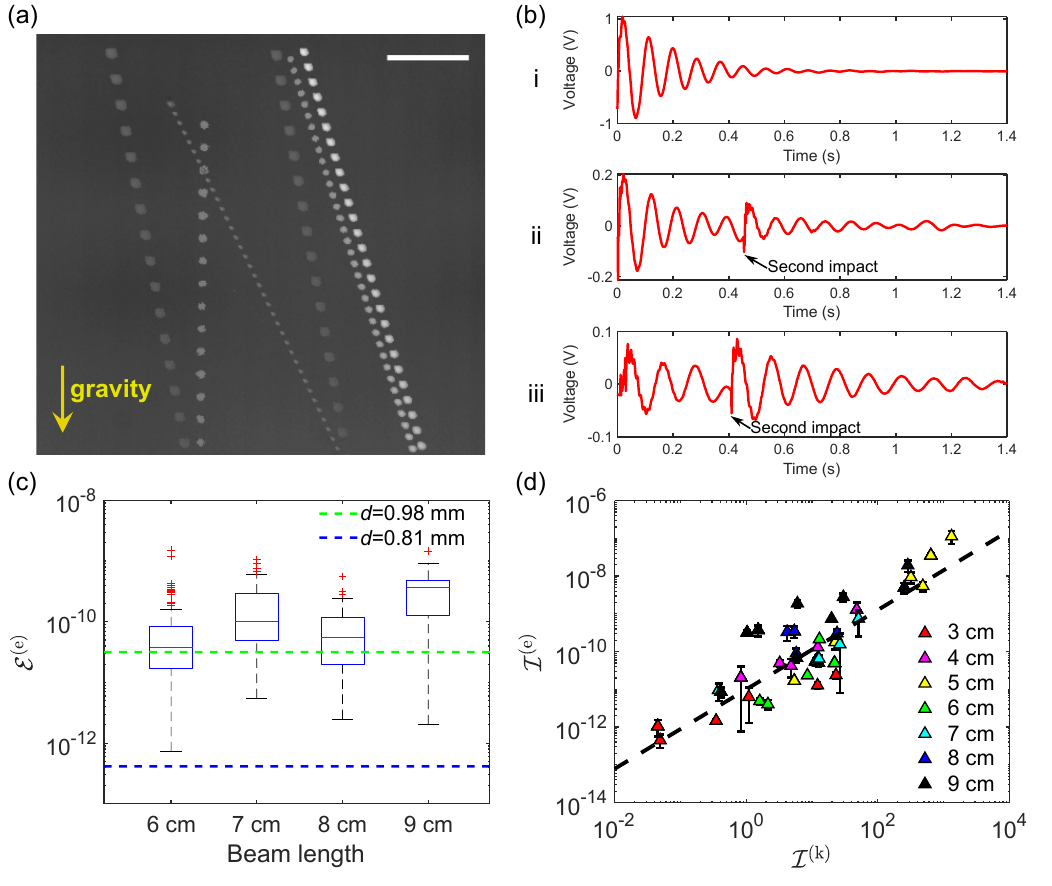}
\caption{(a) Overlaid high-speed camera images capturing the trajectories of raindrops at a rainfall rate of 1.3 mm/h. Six individual droplets are observed descending along distinct paths. The frame interval for the trajectory of all droplets is 0.4 milliseconds. (Scale bar = 5 mm) (b) Representative voltage signals recorded during natural rainfall (2.1 mm/h) using a 6 cm beam. The complete dataset is available in Figure S11. Each panel highlights a distinct event to demonstrate the variability in droplet impacts during actual rain. Panel (b-i) shows a damped harmonic oscillation resulting from a single raindrop impact. Panels (b-ii) and (b-iii) exhibit superimposed signals generated by the sequential impact of two droplets. In Panel (b-ii), the earlier-arriving droplet induced a larger amplitude, whereas in Panel (b-iii), the later-arriving droplet produced a stronger response. (c) Time-averaged electrical energy generated by actual raindrops versus beam length. The values represent the average electrical energy, calculated from signals extracted from individual events within the overall electrical signal. For the 6 cm, 7 cm, 8 cm, and 9 cm beams, the signals were obtained during precipitation levels of 0.44–1.67, 3.64–5.53, 2.13, and 0.42–16.47 mm/hr, respectively. The green and blue dashed lines represent the expected electrical energy generated when droplets with diameters of 0.98 mm and 0.81 mm, respectively, impact the tip of the beam at a velocity of 3.01 m/s. (d) Electrical energy intensity versus kinetic energy intensity on a log-log scale. The dashed line is a fitted curve with an $R$-squared value of 0.9.}
\label{Fig_rain}
\end{figure*}

We conducted experiments to observe the response of a piezoelectric beam to raindrop impacts during actual rain. The voltage response of the beams was measured using a device with four parallel connected beams. Through a total of 47 experiments, we evaluated the generation of electrical energy under actual rain conditions. In addition, we measured the velocity and size of actual raindrops using a high-speed camera. Figure~\ref{Fig_rain}(a) illustrates the falling trajectory of the raindrops, and this data was subsequently utilized to calculate the kinetic energy intensity.


Figure 6(b) shows several representative events from the filtered signal with a 6 cm beam. Figure 6(b-i) displays a damped harmonic oscillation caused by a single droplet impact, while Figures 6(b-ii) and (b-iii) show overlapping signals resulting from two temporally adjacent impacts. The amplitudes of the signals also vary significantly. This variability arises from various uncertainties under real rainfall conditions, including changes in droplet size, velocity, and rainfall intensity. Unlike the controlled single-drop experiments, raindrops do not have a uniform size but follow a log-normal distribution (Figure S10). Our data shows that the mean raindrop diameter is 0.98 mm for precipitation rates below 0.5 mm/hr and 0.81 mm for precipitation rates above 0.5 mm/hr. Additionally, raindrops can land randomly on different parts of the beam, impact the beam simultaneously and overlap signals, or create signal gaps when rainfall intensity is low. Naturally, actual rain data contain more complex signals compared to controlled single-drop tests. 

Despite the complexity and randomness of actual rain events, we selected only clearly distinguishable impact signals from real rain data to evaluate whether the findings from single drop tests still valid. We then compared the average electrical energy generated per raindrop impact with the results from single-drop experiments (Figure~\ref{Fig_rain}(c)).

Since natural raindrops are smaller than those generated using a syringe in our controlled experiments, we also estimated the expected electrical energy output for single droplets with a size equal to the mean raindrop diameter we measured. The mean electrical energy generated by the impact of actual rain was on the same order as the expected energy levels produced when droplets with diameters of 0.81 mm and 0.98 mm hit the tip of the beam. Given that only clear signals were considered, if all signals were included, the average harvested energy from actual rain would likely be lower. This confirms that individual impact signal analysis largely aligns with the single-drop test predictions.


Next, to analyze the overall trend of the actual rain data, we found the best-fit curve (Figure~\ref{Fig_rain}(b)). When the data were fitted to a power function, the $R$-squared value was 0.9, with the exponent at 1.05. In contrast, the single-droplet test exhibited a linear trend, suggesting that the actual rain system introduces additional complexities that result in a different energy conversion pattern.

\section{Discussion} 

Inspired by the fluttering motion of leaves under external forces, we developed a rain energy harvesting device using a piezoelectric beam that converts raindrop impact into electrical energy. Through theoretical and experimental analysis, we examined the energy conversion process to understand its mechanisms and optimization potential. Our results demonstrate that the conversion of rainwater kinetic energy into bending energy is length-dependent, while the conversion of bending energy into electrical energy is area-dependent. Ultimately, the total conversion from kinetic energy to electrical energy becomes independent of length when $L\geq5$ cm. This implies that within the tested length range (5–9 cm), the device design can be freely adjusted without affecting energy conversion efficiency.

Despite the significance of this study in exploring energy conversion in cantilevered piezo beams, there are certain limitations and areas for improvement. First, we observed a discrepancy between single drop tests and real rain data. In controlled single drop experiments, energy conversion followed a linear trend, whereas in actual rain conditions, a power function provided a better fit. This difference likely arises from the inherent randomness in rainfall characteristics, including droplet size distribution, impact locations, precipitation intensity, and environmental factors such as wind. Understanding and modeling these stochastic effects would enhance the predictability and efficiency of rain energy harvesting systems.

To evaluate the potential efficiency of raindrop energy harvesting, we compared the energy yield per unit area with that of wind farms. A typical wind farm requires approximately 140 acres to generate 1 megawatt of power \cite{owoe}. This translates to an energy intensity of roughly 64 $\mathrm{kJ/m^2hr}$ when normalized to our experimental unit ($\mathrm{J/m^2hr}$), assuming continuous and standard conversion efficiencies. In contrast, the energy intensity obtained from one single raindrop harvesting experiment at a release height of 20 cm was approximately 2.16 $\mathrm{mJ/m^2 hr}$. By cascading raindrop impacts through stacked piezoelectric beams, it may be possible to amplify the harvested energy per unit ground area, improving spatial efficiency. 750 beams could be placed 20 centimeters apart along a total height of the wind turbine (approximately 150 m tall \cite{DOE}), which could increase energy intensity to 1.62 $\mathrm{J/m^2 hr}$. Furthermore, while our study isolated the contribution of rainfall, real precipitation events are often accompanied by wind. This additional mechanical input could excite the beam structures and contribute supplementary energy, suggesting potential benefits from multi-modal environmental harvesting.

Additionally, we found that residual droplets accumulating on the beam reduced energy conversion efficiency. This suggests that incorporating features to shed droplets such as non-wetting coatings or biomimetic structures like pointed drip tips in plants~\cite{liu2023efficient, wang2020apex} could mitigate this issue and maintain optimal performance. Piezoelectric beams also offer the advantage of being highly adaptable in their structural design. For instance, artificial leaves shaped as drip tips can be assembled into tree-like configurations to form a piezoelectric beam-based power station. Such a design leverages structural benefits while avoiding the habitat loss and ecological disruption commonly associated with solar or wind energy systems. Furthermore, if animals such as squirrels or birds interact with the structure, their movements could naturally contribute to additional energy generation. Alternatively, a funnel structure can be incorporated to collect raindrops and direct them toward the beam tip, thereby minimizing the effects of residual mass and concentrating the impact at a single point. Implementing such design enhancements is expected to significantly improve energy harvesting efficiency beyond the levels observed under the experimental conditions presented in this study.

Beyond rainwater harvesting, our device has broader applications in various environments where droplet impacts occur, such as showers, irrigation systems, or industrial processes. For example, there have been many attempts to design fog-harvesting systems motivated by nature to address freshwater scarcity\cite{kennedy2024bio}. By integrating a rain-energy harvesting system into the droplet collection process of fog harvesting, we may be able to tackle both water and energy shortages simultaneously. Furthermore, it can be integrated with existing renewable energy systems. Attaching a cantilevered beam beneath a solar panel could enable dual-mode energy harvesting—solar energy in dry conditions and piezoelectric energy during rainfall. Given that energy conversion efficiency remains independent of length beyond a certain threshold, this technology can be adapted to diverse deployment scenarios.

By deepening our understanding of the energy conversion process and optimizing structural designs, piezoelectric rain energy harvesters can contribute to the development of more efficient and versatile renewable energy solutions. Future research should focus on refining predictive models for real rain conditions and exploring advanced materials and structures to maximize energy conversion efficiency.

\section{Materials and Methods}

\subsection*{Beam materials}
As a cantilever beam, a thin piezoelectric film, LDT4-028K from TE Connectivity Co. was used. The dimensions of the beam are 2.2 cm in width, 17 cm in length, and $22.9\times10^{-4}$ cm in thickness. According to the manufacturer data sheets, the total thickness is known to be 157 µm and the capacitance is 11 nF. In addition, an important characteristic of the beam is its flexural rigidity ($EI$). The flexural rigidity of the beam was calculated experimentally using a piezoelectric beam of different lengths and a small mass of known value placed at the end of the beam. The deflection of the beam for these measurements was calculated using the MATLAB image processing toolbox. The equation of the measured deflection ($y(x)$) is written as 

\begin{equation}
\label{EI}
    y(x) = \frac{Fx^2}{EI}\left( \frac{L}{2}-\frac{x}{6} \right) 
\end{equation}
where $F$ is the downward force from the small mass, $L$ is the beam length, and $x$ is the position of the small mass from the fixed point.
Through equation (1), flexural rigidity was calculated 40.2 \textmu $\mathrm{Nm^2}$ on average. 



\subsection*{Drop releasing experimental setup}
Figure~\ref{Fig_char}(a) shows the experimental setup and controlled dimensions. A horizontal rail was attached to a vertical stand to allow adjustments to the height and position of the droplet release mechanism. At the end of the rail, we mounted a 15G blunt needle with an outer diameter of 18 mm, which was connected to a syringe via tubing. A syringe pump ensured consistent droplet release at a specified rate. The droplets generated with this setup had an average mass of 29 mg.

\subsection*{Beam's tip displacement measurement}
Droplets were released from heights ranging from 20 to 70 cm, with their horizontal positions adjusted to fall between the free end and the fixed end of the cantilever. The primary beam lengths tested ranged from 3 to 9 cm. All droplet releases were recorded using a high-speed Photron camera (Photron, Tokyo, Japan) equipped with a 50 mm Nikkor lens (Nikon, Tokyo, Japan). Deflection data were tracked from the moment of impact until movement was significantly reduced, typically around 15 seconds. The deflection was analyzed using MATLAB’s image processing toolbox, computed from a point on the edge of the piezoelectric beam and normalized with its initial position.

\subsection*{Voltage measurement} 
The voltage generated by the piezo cantilever beam as it is deformed by the water droplets was recorded using a USB-6002 DAQ (National instruments, Austin, Texas) card. The piezoelectric beam had two electrodes on the terminal end, one for voltage output and one for ground. These outputs were connected to the USB-6002 DAQ card. Using the USB-6002 DAQ the voltage was recorded through a Matlab interface with a 1000 Hz sampling rate. The voltage, like the tip displacement, was recorded until the movement stopped.

\subsection*{Single beam test condition} 
We tested a single beam under various conditions using a droplet releasing mechanism and a voltage recording mechanism. 
The following experiments were conducted: (1) the simplest experiment, in which a single droplet was dropped onto the tip of the beam, (2) a single droplet impacting a midpoint of the beam instead of the tip, and (3) multiple droplets impacting the tip. The experimental conditions for each case are as follows. We started with the simplest model: dropping a single droplet on the beam's tip. The length of the beam was adjusted to 3-9 cm, and the drop height was adjusted to 20-70 cm. Next, to observe droplets falling in the middle of the beam, we adjusted the position of the horizontal rail. The beam length was adjusted to 3-9 cm, and the horizontal rail position was controlled in increments of 1 cm from the tip to (L-1) cm. In this case, the drop release height was fixed at 60 cm. Lastly, an experiment was conducted to investigate the effects of multiple droplets. The speed of the syringe pump was adjusted so that one droplet fell approximately every 850 ms, and 10 droplets were released at the same position.

\subsection*{Field-test device} 
We designed a device to measure voltage in actual rainy conditions in the field by connecting four beams simultaneously (Figure S9). The device consists of two parts: the main body for securing the beams and a lid to protect the wires and electronic components from rain. After assembly, the overall dimensions of the device are 15 cm in width, 25 cm in length, and 12 cm in height. The designed device was printed using Ultimaker S5 (Ultimaker, Utrecht, Netherlands) with PLA Filament (Ultimaker, Utrecht, Netherlands). After printing the device, the beams were secured to it, and the wires were connected in parallel to the USB-6002 to analyze the voltage signals from all four beams simultaneously.

\subsection*{Field tests under natural rainfall} 
We conducted the rain energy harvesting device under actual rain condition outside of Riley-Robb hall in Cornell University (GPS: 42.45, -76.47). We selected a region where less wind is blowing to avoid any effect of wind. To minimize the effect of wind, we also placed our device in the rectangular container. The beam length range from 3 cm to 9 cm with an increment of 1 cm was tested. Each beam length was tested at least 4 times. The obtained voltage data was further processed using high-pass filtering to remove external influences (e.g., wind) that were not caused by raindrop impacts.

While we were collecting the voltage data, we measured precipitation with the three cups placing on top of the device. A range of precipitation was (0.01 - 49.3 mm / hr) measured. 

\subsection*{Actual rain size and velocity} 
To determine the size and falling speed of raindrops from actual rain data, we captured the rain using a high-speed camera. The experiment was carried out outside of Riley-Robb hall in Cornell University (GPS: 42.45, -76.47). We mounted a high-speed Photron camera (Photron, Tokyo, Japan) with a 105 mm Nikkor lens (Nikon, Tokyo, Japan) on the laboratory window and focused it at a specific distance. Near the focal point, we placed three cups to measure rainfall and recorded the average rainfall over 10 to 30 minutes depending on the intensity of the rain. Simultaneously, we captured the raindrops entering the camera's focus at 5000 frames per second. At the different condition of the rain, the size of the raindrops was analyzed using FIJI software, and the speed was analyzed using Tracker software. The actual raindrop size and velocity are later used to calculate the kinetic energy intensity of rain. The detailed calculation process can be found in Supplementary Section E.

\section{Acknowledgement}
\subsection{Funding}
J.Y. and S.J. acknowledge funding support from the National Science Foundation (NSF) grant (CMMI-2042740).

\subsection{Author contributions}
S.J. conceived the original idea for the study. J.Y., A.K., and K.T. conducted the experiments and analyzed the data. J.Y. and S.J. developed the theoretical framework. All authors participated in the preparation and revision of the manuscript.

\subsection{Competing interests}
The authors declare no competing financial interest. 

\subsection{Data and materials availability:}
The STL files for the rain energy harvesting device, and the data and the Matlab code for the main figures are available at https://osf.io/fbu2q/.


%

\end{document}